\documentclass[aps,pre,twocolumn,11pt]{revtex4-1}
\usepackage{graphicx,bm}
\usepackage{amsfonts,amsmath}

\begin{document}

\title{Centripetal focusing of gyrotactic phytoplankton in solid-body rotation}

\author{M. Cencini$^{1}$, M. Franchino$^{2}$, F. Santamaria$^{3}$ and G. Boffetta$^{3}$}
\affiliation{
$^1$Istituto dei Sistemi Complessi-CNR, via dei Taurini 19, 00185 Rome, Italy \\
$^2$Department of Life Sciences and Systems Biology,
University of Torino, via Accademia Albertina 13, 10123 Torino, Italy \\
$^3$Department of Physics and INFN, 
University of Torino, via P.Giuria 1, 10125 Torino, Italy
}
\date{\today}

\begin{abstract}
A suspension of gyrotactic microalgae \textit{Chlamydomonas
augustae} swimming in a cylindrical water vessel in solid-body
rotation is studied.  Our experiments show that swimming algae form
an aggregate around the axis of rotation, whose intensity increases with
the rotation speed. 
We explain this phenomenon by the centripetal orientation of the 
swimming direction towards the axis of rotation.
This {\it centripetal focusing} is contrasted by diffusive fluxes
due to stochastic reorientation of the cells. 
The competition of the two effects lead to
a stationary distribution, which we analytically derive from a refined
mathematical model of gyrotactic swimmers. The temporal evolution of
the cell distribution, obtained via numerical simulations of the
stochastic model, is in quantitative agreement with the experimental
measurements in the range of parameters explored.
\end{abstract}

\maketitle
\section{Introduction\label{sec:1}}

Many phytoplankton species are able to swim and their motility affects
several basic processes in their life and ecology
\cite{Smayda97,reynolds2006,Kiorboe,elgeti2015}.  Swimming allows
phytoplankton to explore the water column, moving from the well-lit
surface layers during the day to the nutrients-rich deeper layers at
night \cite{lieberman1994,cullen1985} (see also \cite{bollens2010} and
references therein). In order to perform this vertical migration, many
phytoplankton cells are guided by an orienting mechanism when
swimming.
One of the simplest mechanism leading to orientation in the vertical 
direction is bottom-heaviness \cite{wager1910}.
The unbalanced distribution of mass inside the cell induces a mechanical
torque, due to gravity and buoyancy forces, which orients the cell in
the direction opposite to gravity. This torque competes with the 
viscous torque, due to the hydrodynamic shear, which tends to rotate the 
cell. 
When the swimming direction results from the balance between this two
torques, the organism is said to be {\it gyrotactic}
\cite{Kessler1985,Pedley1987,Pedley1992}.

Gyrotaxis generates remarkable spatial distributions of swimming
cells.  In laminar conditions, it produces beam-like accumulations in 
vertical pipe flows \cite{Kessler1985} and concentrated thin layers 
in horizontal shear flows \cite{Durham2009}.
More recently, it has been shown by theoretical analysis and
numerical simulations, that gyrotaxis induces intense small-scale fractal
clustering in turbulent flows 
\cite{Durham2013,zhan2014,fouxon2015,gustavsson2015}, 
providing a mechanisms for the micro-patchiness observed
in motile phytoplankton \cite{malkiel1999}.
Turbulence is characterized by extreme fluctuations of fluid 
acceleration, which can locally exceed gravity even at moderate Reynolds 
numbers \cite{LaPorta2001}. 
The origin of these fluctuations has been traced back to small-scale
vortices generating intense centripetal accelerations \cite{Biferale2005}. 
For applications in turbulent flow, the gyrotactic model has been recently
modified to take into account the effect of fluid acceleration in the 
mechanical torque \cite{JOT}. Remarkably, numerical simulations
of turbulent flows have shown that the effect of fluid acceleration is to 
enhance the patchiness of gyrotactic phytoplankton by gathering cells
into small-scale vortices \cite{DCDBSCB14}.

\begin{figure}[htb!]
\begin{minipage}[c]{0.49\columnwidth}
\includegraphics[width=\columnwidth]{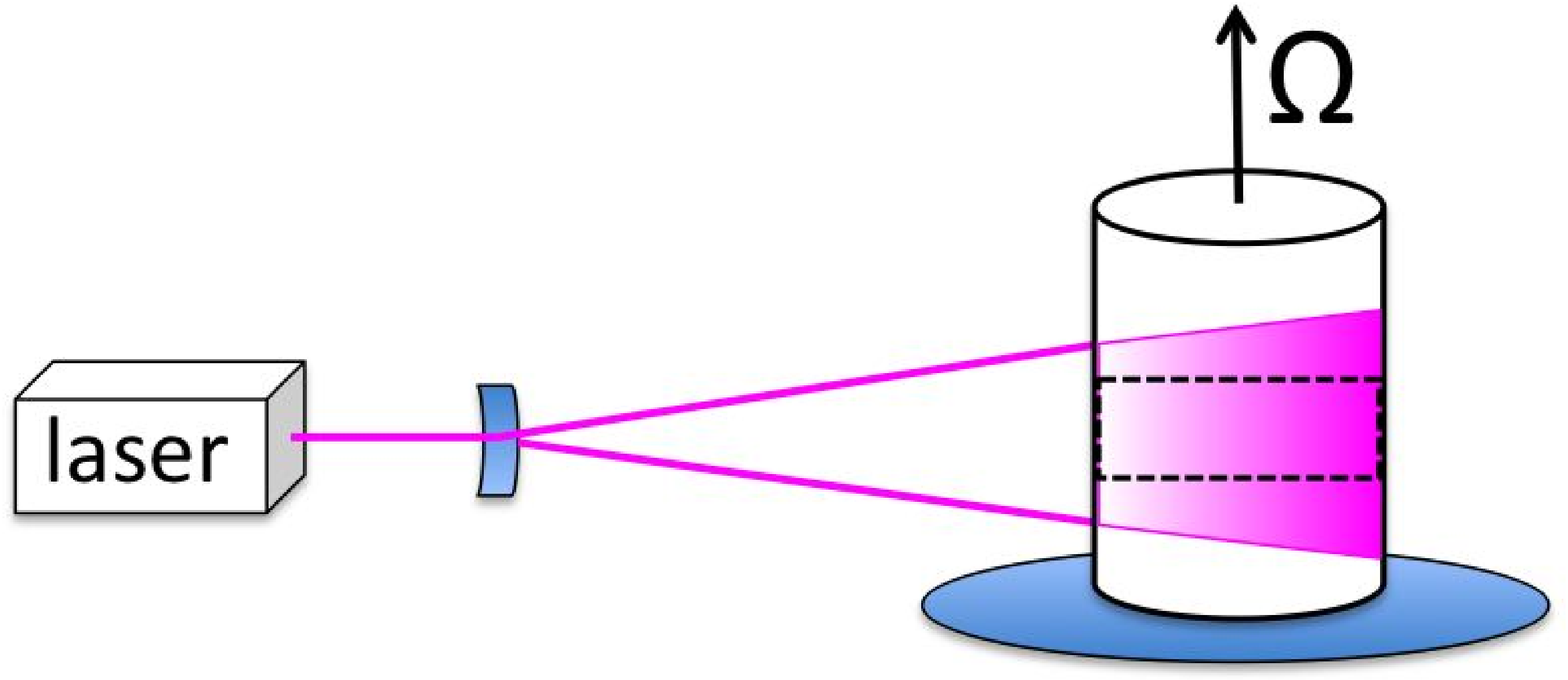}
\end{minipage}
\begin{minipage}[c]{0.49\columnwidth}
\includegraphics[width=\columnwidth]{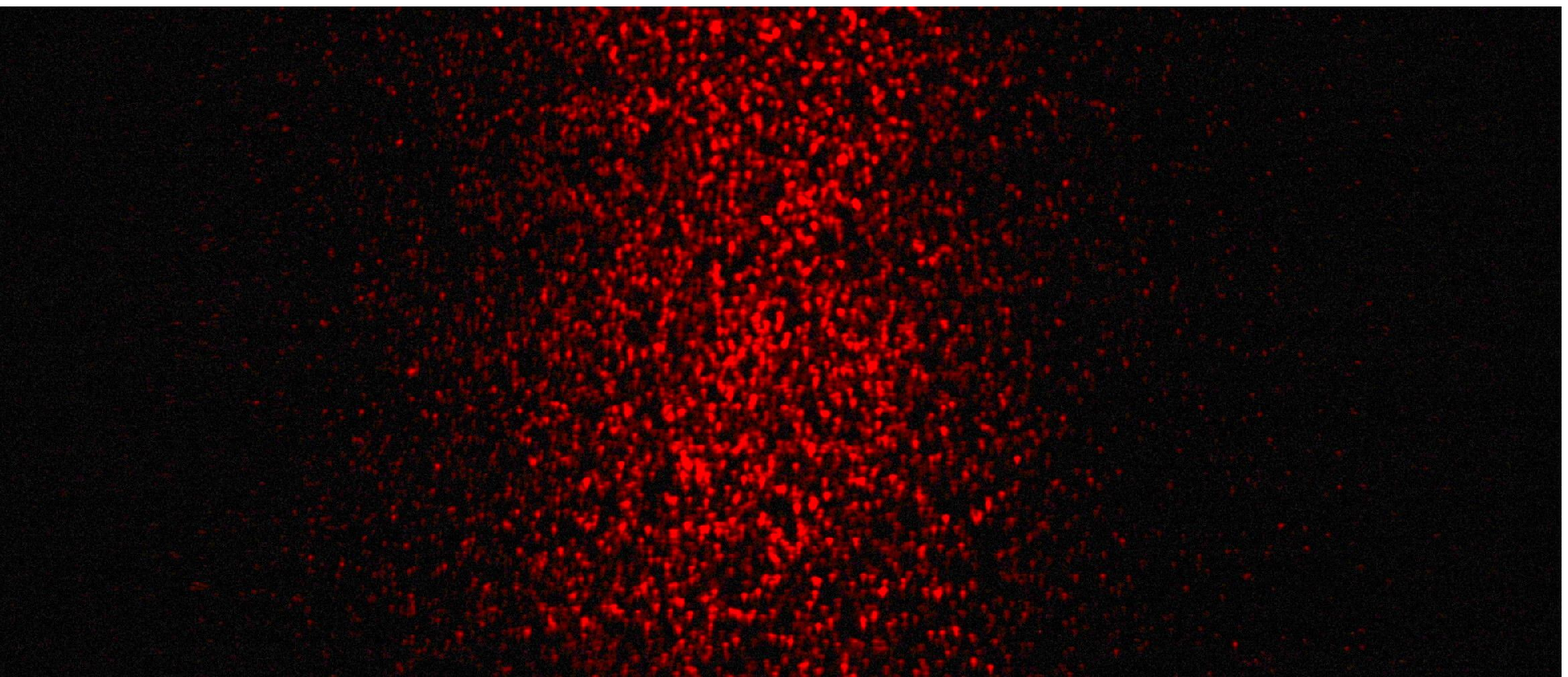}
\includegraphics[width=\columnwidth]{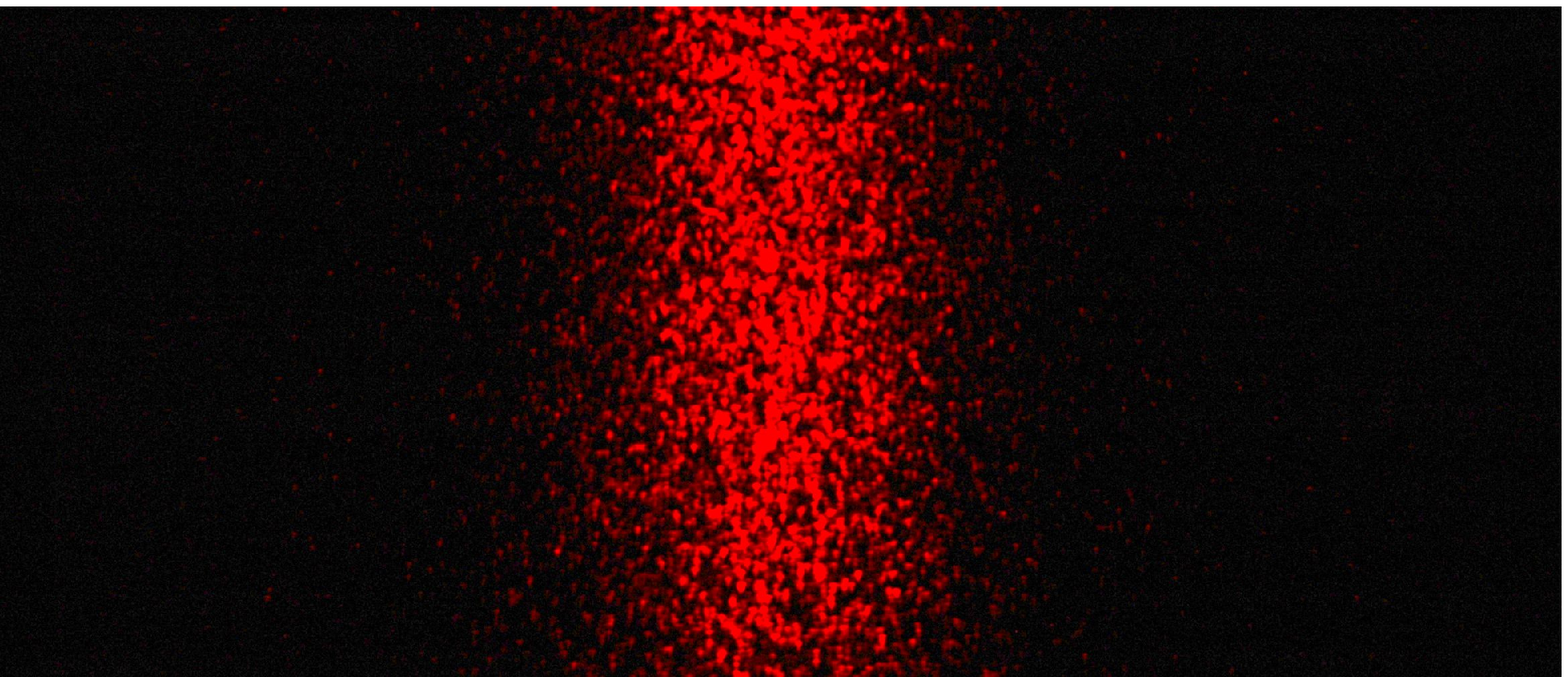}
\end{minipage}
\caption{Online version in colour.
Sketch of the experimental setup. The cylindrical vessel
filled with a suspension of {\it C. augustae} is placed over a table
rotating with angular velocity $\Omega=2 \pi f$.  A blue laser
(power $100 \, mW$, wavelength $\lambda=450 \, nm$) vertical sheet,
generated by a cylindrical lens, is directed in the central plane of
the vessel. Fluorescent images of cells are acquired by a Nikon CCD
camera at resolution $3000 \times 2000$ pixels with a low-pass red filter
at $600 \, nm$.
The rectangular area represents the measurement region.
The pictures on the right are two examples of the images (central part)
taken by the camera at the time $t=600\, s$ for $f=5\, Hz$ (top) and 
$f=8\, Hz$ (bottom).}
\label{fig1}
\end{figure}

In this work, we study experimentally, analytically and numerically
the behavior of gyrotactic swimming cells in a controlled environment
of uniform vorticity -- i.e. a cylindrical water vessel in solid-body
rotation.  Preliminary, qualitative, results of the experiment were
described in Ref~\cite{DCDBSCB14}. We show that phytoplankton cells,
initially uniformly distributed in the container, aggregate towards
the axis of rotation eventually reaching a stationary, Gaussian-like 
distribution.
We show analytically and by means of numerical simulations that a
stochastic formulation of the refined gyrotactic model \cite{DCDBSCB14},
which takes into account the fluctuations of the swimming direction,
is able to reproduce quantitatively the time evolution of a distribution of 
cells and the asymptotic accumulation at different rotation frequency.

The remaining of the paper is organized as follows. In
Section~\ref{sec:2} we introduce the experimental method, the
theoretical model, the different analytical
techniques and the numerical method.  In Section~\ref{sec:3} we
discuss the data analysis and we compare the experimental results with
the theoretical and numerical predictions.  Finally,
Section~\ref{sec:4} is devoted to conclusions. 
Technical details on the analytical computations are presented 
in the Appendices.

\section{Methods}
\label{sec:2}

\subsection{Laboratory experiments}
\label{sec:2.1}
We performed laboratory experiments using cultures of the unicellular
freshwater flagellate strain of Chlorophyceae, {\it Chlamydomonas
augustae}.
{\it C. augustae} is
inoculated in $250 \, ml$ Erlenmeyer flasks containing $200 \, ml$ of
liquid Z medium \cite{andersen2005}.  Cultures are kept at constant
temperature of $25^{o} \, C$ under artificial illumination provided by
fluorescent light producing about $2000 \,lux$ with a $16:8 \, h$
light:dark cycle.  Culture growth is estimated by measuring the dry
biomass concentration and the number of cells. Samples for dry weight
($dw$) calculation are taken in triplicate and a gravimetric
determination is performed \cite{cz_00}.  Triplicate
cell counts are carried out for each sample by loading
$0.01 \, ml$ of sample on a Thoma's counting chamber
and the averaged value was determined.  Cell counts are performed by
using a hemocytometer.  Experiments are realized
about $3$ weeks after inoculation of the culture and after dilution
into fresh medium to reach a concentration of 
$\simeq 5 \times 10^4 \, cell/ml$ to avoid collective phenomena.  

The experimental setup is sketched in Fig.~\ref{fig1}.  A suspension
of gyrotactic cells is placed in a small cylindrical vessel of inner radius
$R=18 \, mm$ and height $H=57 \, mm$ over a plate that
rotates at angular velocity $\Omega$, digitally controlled.
After a short spin-up time of order
$R/(\nu \Omega)^{1/2}$ ($\nu$ is the kinematic viscosity of water)
the fluid in the vessel reaches the solid-body rotation regime 
\cite{gh_jfm63}, at which we defines the reference zero time for the
experiment. We remark that this regime is reached after a series of
physical processes characterized by complex secondary flows. 
Nonetheless, this initial transient is negligible since the spin-up 
time for our vessel is of the order of few seconds and the
displacement of the swimming cells in this transient is very small.

All these preparatory processes are performed in darkness or with a
low power red light at wavelength $655 \, nm$ not seen by algae
\cite{harris2009}.
At time $t=0$ and every $\Delta t$ (typically
$\Delta t=15 \,s$) a blue laser sheet (power $100 \, mW$, wavelength
$\lambda=450 \, nm$) from one side of the vessel is turned on for less
than one second and a picture of the vessel is taken by the camera 
and acquired by the computer.  
Each experiment, for a specific value of $\Omega$, lasts for $15$
minutes after which the plate is stopped and the culture is homogenized.  
Control experiments are performed in similar conditions by using a suspension
of cells killed with a solution of $8 \%$ v/v ethanol.  The
frequency of rotation $f=\Omega/(2 \pi)$ ranges between $4 \, Hz$ and
$8 \, Hz$, corresponding to a centripetal acceleration $a_c=\Omega^2 R$
at the border of the vessel between $11 \, m/s^2$ and 
$45 \, m/s^2$, in all cases larger than the gravitational acceleration.

The laser sheet induces fluorescence of the {\it C. augustae} cells whose
emission spectrum has a peak around $685 \, nm$ (determined by a
spectrophotometer) \cite{harris2009}.  Fluorescence images are taken
by a digital camera (resolution $3000 \times 2000$ pixels) with a red
filter (B+W $091$ Dark Red) which cuts all wavelengths below $600 \,
nm$. 
Samples of the fluorescence images are shown in Fig.~\ref{fig1}.
Calibration of the images with the cell density is
performed by taking pictures of homogeneous suspension (not in
rotation) with known concentrations (determined by a hemocytometer).  
Images of the vessel filled with distilled water are used to define the
background noise (due to background light and CCD noise) which is averaged
over realizations and space and removed from the experimental images.
Spatial calibration is performed by a micrometric pattern inside the 
vessel filled with water, which is also used to compensate the optical 
deformation induced by the cylindrical surface of the vessel.

Images acquired by the camera at different times are used to measure
the evolution of the radial density $n_{exp}(r,t)$ of algae, after
averaging over the vertical direction (we do not observe significant
dependence of the local density on the vertical coordinate).  The
radial density is computed over a portion of the image of height
$h=30$ mm and width $2R_v=24$ mm centered at the cylinder
axis. 
We limit the acquisition to $R_v<R$ in order to reduce optical 
deformations and possible wall effects.

\subsection{Mathematical models}
\label{sec:2.2}
The mathematical model for gyrotactic algae was introduced by
Kessler and Pedley \cite{Kessler1985,Pedley1987,pk_jfm90},
on the basis of the observation that bottom-heavy swimming micro-organisms 
focus in the center of a pipe when the fluid flows downwards. 
The swimming direction results from
the competition between gravity-buoyant torque, due to
bottom-heaviness, and the shear-induced viscous torque. Recently, the
model has been extended to include the acceleration induced by the
fluid flow \cite{JOT,DCDBSCB14}, an effect which can be important in
turbulence \cite{LaPorta2001}.

Due to the small size ($\sim 10\mu m$) and the small density mismatch
with the fluid (less than $5\%$), cells are represented as point-like,
spherical and neutrally buoyant particles (for a recent refined
analysis see Ref.~\cite{Omalley2012}) transported by the fluid
velocity ${\bm u}({\bm x},t)$ at cell position $\bm x$ with a
superimposed swimming velocity ${\bm v}_s$
\begin{equation}
\dot{\bm x} = {\bm u} + v_s {\bf p}\,.
\label{eq:2}
\end{equation}
The magnitude of the swimming velocity, $v_s$, is assumed constant,
while the swimming direction ${\bf p}$ evolves
as
\begin{equation}
\dot{\bf p} = - \frac{1}{2v_o}\left[\bm A - (\bm A \cdot {\bf p}) 
{\bf p}\right] 
+\frac{1}{2} {\bm \omega} \times {\bf p} +{\bf \Gamma_{\mathrm r}}\,.
\label{eq:3}
\end{equation}
The first term describes the orientation towards the direction
opposite to the acceleration $\bm A$ in the cell reference of frame
with characteristic speed $v_o=3 \nu/\delta$ ($\delta$ is the cell
center-of-mass displacement relative to the geometrical center and
$\nu$ the fluid kinematic viscosity). In the original model only
gravity is taken into account and ${\bm A}={\bm g}$. Here, following
\cite{JOT,DCDBSCB14}, we consider the total acceleration, due to
gravity and fluid acceleration ${\bm a}=d {\bm u}/ d t$ (acceleration
due to swimming is neglected), so that $\bm A=\bm g-\bm a$.  The
second term in the rhs of (\ref{eq:3}) represents cell rotation
due to the local vorticity ${\bm \omega}={\bm \nabla} \times {\bm u}$.
The last term $\bm \Gamma_{\mathrm r}$ represents
rotational diffusion in the swimming directions, which phenomenologically 
models the
intrinsic stochasticity in the swimming behavior \cite{hill97}.

We emphasize the simplicity of the above model that neglects many 
details, including the unsteadiness of swimming due to flagella beating, 
deviations from spherical shape, cell-cell interactions and the feedback 
of cell motion on the surrounding fluid.

\subsection{Deterministic motion in solid body rotation}
\label{sec:2.3}
We study analytically and numerically the motion of gyrotactic
swimmers in the velocity field generated by the solid body rotation of
the cylinder along its vertical axis: ${\bm u}=(-\Omega y,\Omega
x,0)$.  Vorticity is parallel to gravity, ${\bm \omega}={\bm \nabla}
\times {\bm u}=(0,0,2 \Omega)$, and the (centripetal) acceleration
${\bm a}=(-\Omega^2 x, -\Omega^2 y, 0)$ is orthogonal to gravity ${\bm
  g}=(0,0,-g)$.

We start by considering the dynamics in the absence of stochastic
terms (i.e. $\bm \Gamma_{\mathrm r}=0$ in (\ref{eq:3})) whose effects
will be discussed in the following Section. In this limit it is
possible to derive analytically the motion of swimming cells and show
that they accumulate exponentially in time on the axis of rotation.

By introducing cylindrical coordinates with 
${\bf x}=({\bf r},z)$ and ${\bf p}=({\bf p}_r,p_z)$,
such that $\bm A=(\Omega^2 \bm r,-g)$ and $\bm u=(\Omega
\bm r^{\perp},0)$ (with ${\bm r}^{\perp}=(-y,x)$), Eqs.
(\ref{eq:2}-\ref{eq:3}) become
\begin{eqnarray}
\dot{\bf r} &=& \Omega {\bf r}^{\perp} + v_s {\bf p}_r \label{eq:4}\\
\dot{z} &=& v_s  p_z \label{eq:5}\\
\dot{\bf p}_r &=& - \frac{1}{2 B} \left[
\gamma {\bf r} + p_z {\bf p}_r - \gamma ({\bf r} \cdot {\bf p}_r) {\bf p}_r
\right] + \Omega {\bf p}_r^{\perp} 
\label{eq:6} \\
\dot{p}_z &=& \frac{1}{2 B} \left[
1 - p_z^2 + \gamma ({\bf r} \cdot {\bf p}_r) p_z
\right] 
\label{eq:7}
\end{eqnarray}
where ${\bf p}_r^{\perp}=(-p_y,p_x)$. The two parameters $B \equiv
v_o/g$ and $1/\gamma \equiv g/\Omega^2$ represents the characteristic
time of reorientation under gravity and the radial distance at which
fluid and gravitational acceleration are equal.

A solution of the above equations can be obtained under the
hypothesis of local equilibrium in the swimming direction, 
i.e. by assuming $\dot{\bf p}=0$ locally. 
Physically this requires that the characteristic orientation
time $B$ is faster than the typical displacement time.
Within this approximation one can show 
(see Appendix~\ref{app:deterministic}) that
the equilibrium swimming direction ${\bf p}^{eq}$ is simply opposite
to the total acceleration, $-\bm A$, i.e.
\begin{equation}
{\bf p}^{eq}= \frac{\bm a\!-\!\bm g}{|(\bm
a\!-\!\bm g)|}=\left(\frac{-\gamma \bm r}{\sqrt{1+(\gamma r)^2}},\frac{1}{\sqrt{1+(\gamma r)^2}}\right) \,.
\label{eq:8}
\end{equation}
When $\gamma r \ll 1$, the distance from the axis evolves as
$\dot{r}=-\gamma v_s r$ implying that cells position moves
exponentially towards the rotation axis $r=0$:
\begin{equation}
r(t) = r(0) e^{- \gamma v_s t}\,.
\label{eq:9}
\end{equation}
For general values of $\gamma r$ the time evolution of $r(t)$ is
more complicated but asymptotically the above results remain
valid (see Appendix~\ref{app:deterministic} for details).

\subsection{Stationary distribution in presence of rotational diffusivity}
\label{sec:2.4}
The deterministic model predicts that swimming cells should converge
on the rotation axis.  In reality,
stochastic reorientation of the swimming direction, due to
different biological behaviors, causes the cells to weakly deviate
from the convergent trajectories predicted by (\ref{eq:9}), preventing
the population from collapsing onto the rotation axis and eventually
producing accumulation in a finite volume.

Because we are considering stochastic effects, which can be
conveniently modeled as rotational diffusivity \cite{hill97,Pedley1992},
in the following we will not discuss individual trajectories, rather we 
shall focus on the cell density $n(\bm x,t)$ which can be directly compared 
with the experimental measurements.
We report here the basic ingredients and results of our approach,
which is based on the so-called Generalized Taylor Dispersion theory
\cite{brenner1989}, applied to gyrotactic phytoplankton 
\cite{bearon2011,bearon2012}. The interested reader can find details of 
this approach in Refs.~\cite{hill2002,manela2003,bearon2011,bearon2012} 
(see also Appendix~\ref{app:GTD}).

We introduce the probability density function 
$\mathcal{P}({\bm x},{\bf p},t)$ to
find a cell at position ${\bf x}$ swimming in direction ${\bf p}$.
The evolution of $\mathcal{P}$ is ruled by the Fokker-Planck equation
\begin{equation}
\partial_t \mathcal{P} + {\bm \nabla}_{\bm x} \cdot (\dot{\bm x} \mathcal{P}) +
{\bm \nabla}_{\bf p} \cdot (\dot{\bf p} \mathcal{P} - d_{\mathrm r} {\bm \nabla}_{\bf p} \mathcal{P})=0\,,
\label{eq:10}
\end{equation}
where $\dot{\bm x}$ and $\dot{\bf p}$ are given by (\ref{eq:2}) and
(\ref{eq:3}) and $d_{\mathrm r}$ is the rotational diffusivity
coefficient \cite{pk_jfm90,Pedley1992}.  The quantity we are
interested in is the population density, $n(\bm
x,t)=\int \mathrm{d}{\bf p} \,\mathcal{P}({\bm x},{\bf p},t)$.
Essentially, we seek for an effective evolution equation for the
population density in terms of the advection-diffusion equation
\cite{pk_jfm90,Pedley1992,Bees1998,hill2002,manela2003,bearon2011,bearon2012}
\begin{equation}
\partial_t n + {\bm \nabla}_{\bm x} \cdot (\bm V n - \mathbb{D} \bm \nabla_{\bm x} n)=0\,,
\label{eq:11}
\end{equation}
where $\bm V$ and $\mathbb{D}$ are the effective drift and
diffusivity tensor, respectively.  

Leaving the details in the Appendix~\ref{app:GTD}, the main idea for
going from (\ref{eq:10}) to (\ref{eq:11}) is to assume that the
orientation dynamics is the fastest process. Consequently, we can
treat ${\bf p}$ as a random unit vector with probability density
function $f({\bf p})$ given by the stationary solution of the
Fokker-Planck equation $\partial_t f+\bm \nabla_{\bf p} \cdot(
\dot{{\bf p}}f-d_{\mathrm r} \nabla_{\bf p}f)=0$
\cite{pk_jfm90,Pedley1992,Bees1998}.  Within this
approximation the effective drift becomes
$\bm V = \bm u+v_s\langle {\bf p} \rangle= \bm u+v_s \int
\mathrm{d}{\bf p}\,{\bf p}f({\bf p})$.  The derivation of the
diffusivity tensor is more complicated.
The Generalized Taylor dispersion has been used to derive 
$\mathbb{D}$ in homogeneous shear flow \cite{hill2002,manela2003}.
The method can be extended to inhomogeneous shears provided that 
the relaxation in the swimming direction is sufficiently fast 
\cite{bearon2011}.  We are here in a similar situation: 
the equilibrium orientation direction depends on the
distance from the axis, according to (\ref{eq:8}).
We therefore assume that, at each distance $r$, the distribution
of orientations relaxes to a stationary distribution which parametrically
depends  on $r$.  Within this adiabatic approximation, we are able to obtain
explicit expressions for the drift coefficient and the
diffusivity tensor.

Once ${\bf V}(r)$ and $\mathbb{D}(r)$ are known, we can solve (\ref{eq:11}) at
stationarity, i.e. when centripetal flux, controlled by swimming, is
balanced by the diffusive (centrifugal) flux due to random
reorientation.
The main analytical result of our analysis is the explicit expression 
of the stationary radial population density which, for $(\gamma r) \ll 1$, 
takes the Gaussian form (see Appendix~\ref{app:GTD} for details)
\begin{equation}
n_s(r) = \mathcal{N} \exp\left(- \frac{\gamma r^2}{2 v_s B F_3^2(\lambda)}
\right)
\label{eq:12}
\end{equation}
where $F_3(\lambda)$ a dimensionless function of the parameter
$\lambda=1/(2 B d_r)$, and the coefficient
$\mathcal{N}$ can be written in terms of the total number of cells $N_s$
as $\mathcal{N}=N_s \gamma/(2 \pi H v_s B F_3^2(\lambda))$.

From the distribution (\ref{eq:12}) we obtain the average 
distance of the population from the cylinder axis in stationary conditions
which reads
\begin{equation}
\langle r \rangle_s \equiv \frac{\int_{0}^{R} r n_s(r) dr}
{\int_{0}^{R} n_s(r) dr } 
= \sqrt{\frac{2}{\pi}} \left(\frac{v_sB}{\gamma}\right)^{1/2} F_3(\lambda)\,,
\label{eq:13}
\end{equation}
where the subleading contribution from the upper integration 
extreme $R$ has been neglected. 

\subsection{Population evolution: numerical simulations}
\label{sec:2.5}
The temporal evolution of the cell density
cannot in general be obtained analytically. Therefore, in
order to compare the model with the experimental measurements, we
performed a Monte Carlo simulations of (\ref{eq:10}) directly
simulating stochastic trajectories from Eqs~(\ref{eq:4}, \ref{eq:6},\ref{eq:7})
using a Runge-Kutta fourth order scheme (the vertical dynamics
(\ref{eq:5}) is not needed as we are interested in the radial distribution).
The cell's parameters $v_s$, $B$ and $d_r$ are chosen, among the typical
values given in literature, in order to have a good fit of the stationary
distribution, as described below. 
The parameter $\gamma=\Omega^2/g$ is varied according to the experimental 
rotation frequency, with $g=9.8\,m/s^2$.

The simulation is initialized placing $N_s=10^5$ cells randomly in a
circle of radius $R=18$mm with random swimming orientations uniformly
distributed on the unit sphere. We use a time step of $10^{-3}s$ and
store the position and velocities of the cells every $15\,s$, as in
the experiment. From the stored position we can reconstruct the cell
density and other observable. 
In particular, we will be interested in the evolution of the average
radial distance $\langle r(t)\rangle=\int dr n(r,t) r/\int dr n(r,t)$.

\section{Data analysis and results}
\label{sec:3}
In this Section we compare the predictions of the mathematical model
presented in Methods with the outcome of the experiments. The
analytical theory of Sect.~\ref{sec:2.4} provides an explicit
expression for the radial population density $n_s(r)$ at equilibrium, 
while the time evolution of the density is obtained by
stochastic simulations of (\ref{eq:10}).

In order to compare experimental and theoretical data
we have to modify the theory of Section~\ref{sec:2} by adding to the
theoretical population density $n(r,t)$ a constant population of
non-motile cells of uniform density $b$.
Indeed a fraction of the cell population does not move (or it moves 
very slowly) and thus does not contribute to the accumulation.

\begin{figure}[b!]
\includegraphics[width=\columnwidth]{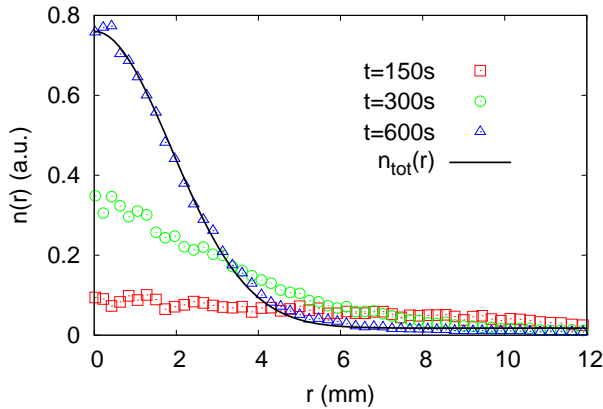}
\caption{Online version in colour.
Evolution of the experimental radial population density 
$n_{exp}(r,t)$ (in arbitrary units) for the experiment at frequency 
$f=7\,Hz$ a time $t=150 \, s$ (red squares), $t=300 \, s$ (green circles) and
at time $t=600 \, s$ (blue triangles) as a function of the distance
from the cylinder axis $r$. 
The black solid line represents the theoretical prediction for the 
total density in stationary conditions 
$n_{tot}(r) = n_s(r) + b$ with $n(r)$ given by (\ref{eq:12}).
The cell's parameters are $v_s=0.1\, mm/s$, $B=7.5\, s$, 
$d_r=0.067\, rad/s$ ($\lambda=1.0$).
The background parameter is fitted to $b=0.017$ which corresponds
to $\beta=N_b/N_s \simeq 1.1$.
}
\label{fig2}
\end{figure}

Therefore, the experimental density $n_{exp}(r)$ will be compared with 
the total density, given by the superposition of the two
populations, $n_{tot}(r,t)=n(r,t)+b$ because, thanks to the dilute
concentration, we can assume that the non-motile population does 
not interfere with the motion of the swimming population.
The total number of cells $N_{tot}$ becomes
\begin{equation}
N_{tot}=N_{s} + N_{b}= 
2 \pi H \int_{0}^{R} r n(r,t) dr + H \pi R^2 b \, .
\label{eq:14}
\end{equation}
By using the total population density, 
the stationary average distance (\ref{eq:13}) becomes
\begin{equation}
\langle r \rangle = \frac{1 + c^2 \beta}{\langle r \rangle_s^{-1} + 
2 c \beta  R^{-1}}
\label{eq:15}
\end{equation}
where $\langle r_s \rangle$ is given by (\ref{eq:13}),
$\beta \equiv N_b/N_s$ is the ratio of the two populations and
$c=R_v/R=2/3$ is a numerical correction due to the fact that
the experimental radial distribution is sampled only up to $R_v<R$.

Figure~\ref{fig2} shows the time evolution of the experimental radial
population distribution for the experiment at $f=7 \, Hz$.  As one can
see, the population of swimming cells progressively concentrates
around the axis of the cylinder ($r=0$).  After the last time shown in
the plot ($t=600 \,s$, which corresponds to the images shown in
Fig.~\ref{fig1}) the distribution remains statistically stationary.  
The theoretical asymptotic distribution (\ref{eq:12})
is used to fit the stationary distribution. 
As in (\ref{eq:12}) the cell's parameters
enter only in the combinations $v_s B$ and $B d_r=\lambda$
we cannot use the stationary distribution to fit all
the parameters.  We have therefore chosen to fix two of the parameters
as given by literature i.e. $v_s=100\,\mu m/s$ and $d_r=0.067\,$rad$/s$
\cite{harris2009,williams2011}, and to use the orientation time $B$
as a fitting parameter.  The resulting value, $B=7.5
\, s$, is compatible with estimations for {\it Chlamydomonas}
\cite{Yoshimura2003,Roberts2006}. 
We remark however that, as clear from the above discussion, other 
combinations of the parameters are possible.
The theoretical asymptotic distribution (\ref{eq:12}), with the 
correction of the background term ($b=0.017$), fits very well the 
experimental data.

\begin{figure}[t!]
\includegraphics[width=\columnwidth]{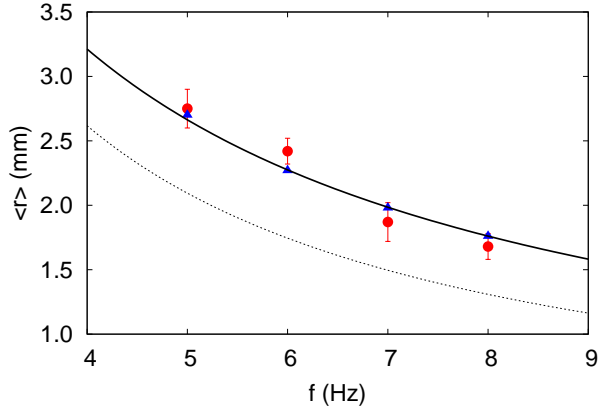}
\caption{Online version in colour.
Stationary radius of cell population, $\langle r \rangle$ from
experimental distribution (\ref{eq:16}) (red circles) and theoretical
prediction, both in the absence of non-motile background cells 
(\ref{eq:13}) (dotted line) and with background contribution 
(\ref{eq:15}) with $\beta=1.1$ (solid line). 
 The error bars on the experimental data are obtained from the temporal
fluctuations of the asymptotic radius for $t>600\, s$.
The blue triangles represent the 
asymptotic values obtained from numerical simulations of a population
of $10^5$ swimmers as explained in Section~\ref{sec:2.5}.
}
\label{fig3}
\end{figure}

\begin{figure}[htb!]
\includegraphics[width=\columnwidth]{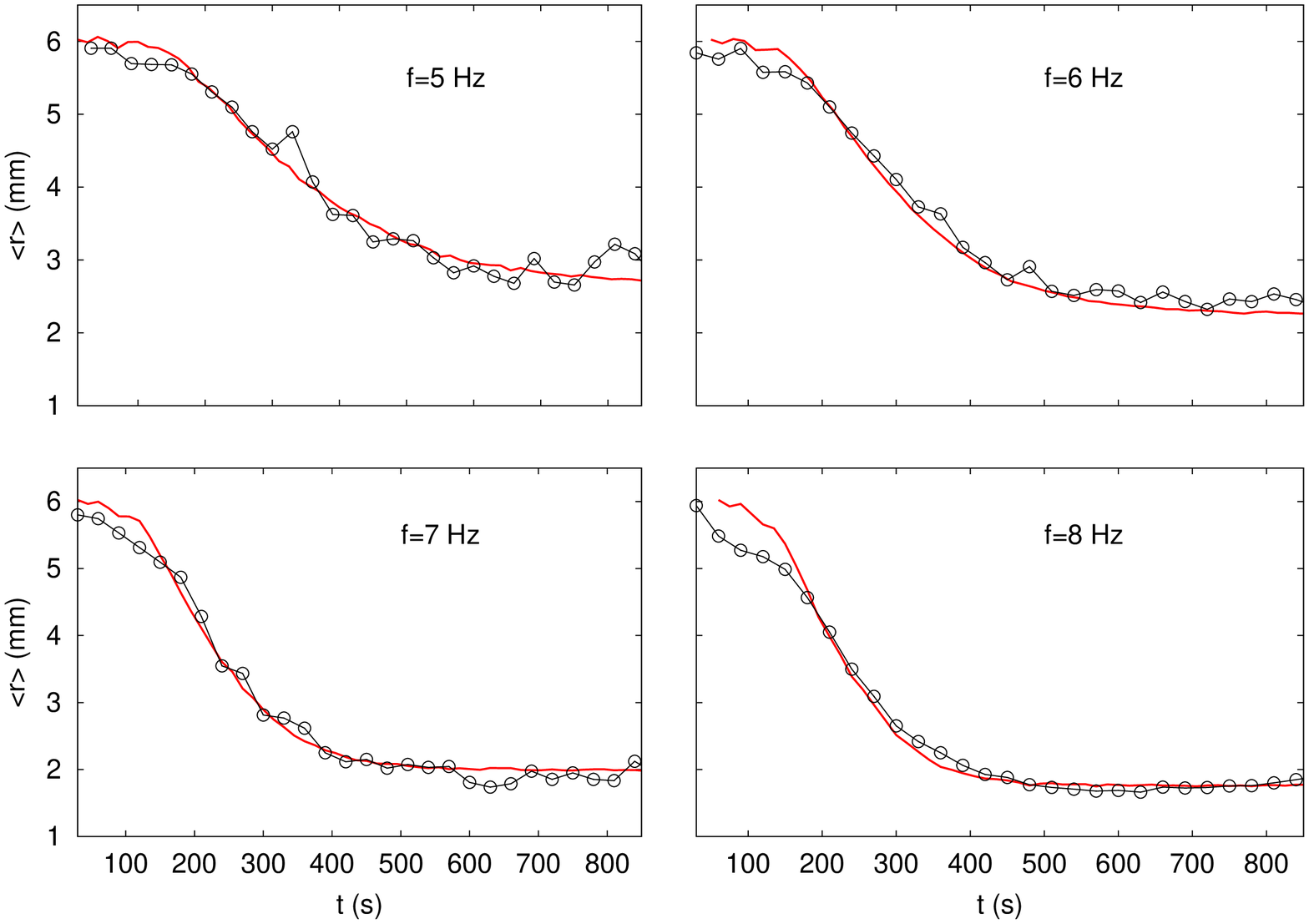}
\caption{Online version in colour.
Cell population radius as a function of time measured in experiments 
(symbols) and in simulations with background $\beta=1.1$ (red solid line).
Model parameters as in Fig.~\ref{fig2}. 
We introduced in the numerical data a time shift $\Delta t$ in the range 
$10-50 \, s$ as the initial phase of spin-up is not reproduced by 
the simulations.}
\label{fig4}
\end{figure}

Figure~\ref{fig3} shows the theoretical and experimental values of the
mean radius $\langle r \rangle$ in stationary conditions.
The experimental stationary radius is obtained by computing, for different
values of the rotation frequency $f$, the time evolution of
\begin{equation}
\langle r(t) \rangle_{exp} = \frac{\int_{0}^{R_v} r n_{exp}(r,t) dr
}{\int_{0}^{R_v} n_{exp}(r,t) dr }
\label{eq:16}
\end{equation}
and looking at the plateau that can be observed at long time values
(see Fig.~\ref{fig4}).  Figure~\ref{fig3} shows both the theoretical
value in the absence of background (\ref{eq:13}), which clearly
underestimates the asymptotic radius, and the expression (\ref{eq:15})
corrected with the background coefficient $\beta=1.1$.  We remark that
the theoretical line in Fig.~\ref{fig3} is obtained without free
parameters, as $\beta$ is fitted from the asymptotic distribution
$n_s(r)$ for a single rotation frequency (Fig.~\ref{fig2}).


By using the numerical method discussed in Section~\ref{sec:2.5} 
it is possible to obtain the time evolution of the radial population density 
$n(r,t)$ and of the mean radius.
We remark that the time evolution of the population density depends 
on a different combination of the parameters with respect to the 
stationary distribution.  Therefore,
the ability to reproduce the experimental dynamics from the numerical
integration of (\ref{eq:4}-\ref{eq:7}), without additional fitting
parameters, is both a test of the validity of the mathematical model
and of the chosen set of parameters.
The time evolution of the mean radius is shown in Fig.~\ref{fig4} for 
the four different values of rotation and with the contribution of the 
background term $b=0.017$ as for the theoretical analysis.
The presence of an initial plateau of
almost constant $\langle r \rangle$ (which is more evident for the
experiments at lower frequencies) is a consequence of the fact that
the average in (\ref{eq:16}) is taken over the inner cylinder of radius 
$R_v < R$ (indeed, the analysis of the numerical data up to $R$ shows that
the plateau disappears). The quality of
the agreement between the theoretical prediction and the experimental
evolution of the mean radius is remarkable.

\section{Conclusion and discussion}
\label{sec:4}
We have studied the centripetal focusing of gyrotactic microalgae
around the axis of a rotating container. By using a refined model of
gyrotactic motion (which takes into account the effect of fluid
acceleration on cell orientation) we derived an effective equation 
for the evolution of
the population density inside the container. The analytical
predictions for the stationary distribution and its characteristic size
reproduce accurately the experimental
data obtained from a population of {\it C. augustae} in a vessel
rotating with different angular velocities.
The time evolution of the distribution is
obtained by stochastic simulations of the Fokker-Planck
equation. Also in this case the results reproduce accurately the
experimental data without additional free parameters.

One of the motivations of our study is to better understand the
behavior of gyrotactic organisms in a turbulent environment (natural
or artificial) characterized by strong region of vorticity. Indeed, 
recent numerical simulations have shown that gyrotactic swimmers are able to
concentrate in regions of high vorticity even if these regions are
very localized and ephemeral, as in the case of homogeneous turbulence
\cite{DCDBSCB14}. 
In this perspective, the vertical solid body rotator has to be considered 
as a ``toy model'' of a turbulent vortex for which analytical
prediction are possible. Similar configurations of uniform vorticity
have been studied recently and proposed for laboratory experiments in 
theoretically controlled conditions \cite{Thorn2010,Pedley2015}.
On the basis of our results, artificial turbulent flows could be 
designed to optimize the trapping of swimming cells in specific regions.

More in general, understanding the interplay between swimming and
fluid transport is crucial to rationalize phytoplankton ecology
\cite{reynolds2006,Kiorboe} and also for industrial applications.
For example, many (motile) microalgae are cultured in photobioreactors to
be commercially used as nutrients, for biofuels production or for
cosmetic industry \cite{borowitzka99}.  As bioreactors work both in
laminar and turbulent fluid motion \cite{bees_interface2013}, 
a better understanding the interaction between fluid motion and 
swimming is fundamental to optimize efficiency.

\begin{acknowledgments}
We thank F. de Lillo for useful discussions and for
carefully reading the manuscript and N. Dibiase for help
with the experimental setup.  We acknowledge the European COST Action
MP1305 ``Flowing Matter'' for support. {\it C. augustae} were provided
by CCALA, Institute of Botany of the AS CR, Tebon Czech Republic.
\end{acknowledgments}

\appendix
\section{Solution of the deterministic model \label{app:deterministic}}

The radial evolution of the deterministic model
(\ref{eq:4})-(\ref{eq:7}) can be solved using the equilibrium
hypothesis (i.e. assuming that $\dot{\bf p}=0$, which should be valid
for times much larger than $B$). The symmetries of the problem suggest
that ${\bf p}_r=\mathrm{p}_r(r) \hat{\bm r}$ and
$\mathrm{p}_z=\mathrm{p}_z(r)$, with
$\mathrm{p}^2_r(r)+\mathrm{p}^2_z(r)=1$. To find the expressions for
$\mathrm{p}_r$ and $\mathrm{p}_z$, it is convenient to multiply
(\ref{eq:6}) by ${\bf p}_r$, using ${\bf p}_r^\perp\cdot {\bf p}_r=0$,
one obtains ${\bf p}_r=-(\gamma \bm
r)/\sqrt{1+(\gamma r)^2}$ and $\mathrm{p}_z=1/\sqrt{1+(\gamma r)^2}$
whose physical interpretation is transparent: ${\bf p}$ aligns in the
direction opposite to the total acceleration (see Eq.~(\ref{eq:8})).

Multiplying (\ref{eq:4}) by $\bm r$ we obtain that the radial distance
evolves according to
\begin{equation}
\dot{r} = -\gamma v_s \frac{r}{\sqrt{1+(\gamma r)^2}}\,.
\label{eq:rmoddot}
\end{equation}
When $\gamma r\ll 1$, Eq.~(\ref{eq:rmoddot}) reduces to
$\dot{r}=-\gamma v_s r$ implying an exponential decay of the distance
from the axis, see Eq.~(\ref{eq:9}).  When $\gamma r \gg 1$,
Eq.~(\ref{eq:rmoddot}) reduces to $\dot{r}\approx -v_s$ so that we have
$r(t)=r(0)-v_s t$, i.e. a linear decrease of the distance from the
axis of rotation.  For generic $\gamma r$, Eq.~(\ref{eq:rmoddot}) is
solved by
\begin{equation}
\frac{r}{r_0} 
\frac{1+\sqrt{1+(\gamma r_0)^2}}{1+\sqrt{1+(\gamma r)^2}}
e^{\sqrt{1+(\gamma r)^2}-\sqrt{1+(\gamma r_0)^2}} = e^{-\gamma v_s t}
\label{eq:solution}
\end{equation}
with $r_0=r(0)$ and $r=r(t)$.  In typical experimental condition we
have $\gamma\approx 65-250$ for $f=4-8$Hz so that for $r_0=O(1)$cm we
have $(\gamma r) \approx 0.6-2.5$. Therefore, at least at the
beginning, there will be deviation from the exponential regime, which
shows up in the latest stage of the (deterministic) evolution.

\section{Derivation of stationary population density in the presence of rotational diffusivity\label{app:GTD}}

Here, we derive the effective drift and diffusion tensor of the
advection diffusion equation (\ref{eq:11}). Then we compute the
stationary population density and use it to derive the average
distance from the cylinder axis.  We detail the
method in two dimensions ($d=2$), because algebra is straightforward
and closed expressions for the quantities of interest can be found.
We then generalize the result to $d=3$, which is the
case relevant for the experiment.
 
\subsection{Analytical approximation in $d=2$\label{suapp:2d}}
As clear from Eq.~(\ref{eq:4}), the fluid velocity causes only
rotation around the cylinder axis and does not contribute directly to
the radial evolution, which is controlled by swimming.  However,
rotation and, in particular, the associated centripetal acceleration
makes the swimming direction to depend on the radial distance, as
clear from Eqs.~(\ref{eq:6}) and (\ref{eq:4}). We can thus study the
problem in two dimensions considering the following dynamics
\begin{eqnarray}
\dot{r} &=& v_s \mathrm{p}_r=v_s \sin\theta\\ \dot{z} &=& v_s \mathrm{p}_z=v_s
\cos\theta\\ \dot{{\bf p}} &=& -{\scriptstyle \frac{1}{2B}} (\bm A(r)-(\bm
A(r)\cdot{\bf p}){\bf p}) +\bm \Gamma_{\mathrm r}\,, 
\label{eq2d:p}
\end{eqnarray}
with $\bm A(r)=\gamma r \hat{\bm{x}}-\hat{\bm{y}}$.
The stochastic term $\bm \Gamma_{\mathrm r}$ represents rotational
diffusion used to model stochasticity in the swimming orientation
\cite{Pedley1987,Pedley1992}. By denoting with $r$ and $z$ the $x-$
and $y-$ components of the position vector, respectively, by defining
$\theta$ as the angle with the vertical and by posing
$\gamma=\Omega^2/g$, we will end up with expressions that can be
directly used in the $d=3$ case.  In $d=2$, Eq.~(\ref{eq2d:p}) can be
conveniently rewritten as a stochastic differential equation for the angle
\begin{equation}
\dot{\theta} = -\frac{1}{2B} (\sin\theta+\gamma r \cos\theta) + \sqrt{2d_{\mathrm r}} \eta(t)\,.
\label{eq2d:thetadot}
\end{equation}
Rotational diffusion simply becomes diffusion of the angle, where
$\eta$ is white noise of unit variance.  By nondimensionalizing the
time one easily derives that the orientation distribution is a
function of $\gamma$ and of the non-dimensional parameter
$\lambda=1/(2Bd_{\mathrm r})$, quantifying the cell stability with
respect to rotational Brownian motion, a large (small) value of
$\lambda$ means that cell orientation is dominated by the bias
(rotational diffusion).

For $\gamma=0$, the model describes the evolution of gyrotactic
swimmers in a two dimensionional still fluid, a problem for which the
effective advection-diffusion equation (\ref{eq:11}) for the
population density has been derived analytically \cite{bearon2011}.
Here, we summarize the results and formulas
needed in the following (see Appendix A.1 of Ref.~\cite{bearon2011}
for a detailed derivation).  The stationary distribution of the
swimming orientation, i.e. the PDF of $\theta$, is the von Mises
probability density function with zero mean, meaning that the average
swimming direction is along the vertical,
\begin{equation}
f^{(0)}(\theta)= \frac{1}{2\pi I_0(\lambda)} e^{\lambda \cos\theta}\,,
\label{eq2d:pdftheta}
\end{equation}
where $I_k(\lambda)$ is the modified Bessel function of the first kind
and order $k$ \cite{mardia}.  In the above equation and in the
following we shall use the superscript $^{(0)}$ to denote the
$\gamma=0$ results. The population density $n(r,z,t)$ is described by
an effective advection-diffusion equation like (\ref{eq:11}) with
drift given by
\begin{equation}
\bm V^{(0)}=\!v_s \int\! d\theta (\sin\theta,\cos\theta) f^{(0)}(\theta) \!=v_s \left(0, \frac{I_1(\lambda)}{I_0(\lambda)}\right)\,,
\end{equation}
and diffusion tensor
\begin{equation}
\mathbb{D}^{(0)}=\frac{v_s^2}{d_{\mathrm r}}\mathrm{Diag}\{D^{(0)}_\perp(\lambda), D^{(0)}_\parallel(\lambda)\}\,,
\end{equation}
where $v_s^2/d_{\mathrm r}$ gives the dimensional contribution;
$\perp$ and $\parallel$ label the direction perpendicular and parallel
to the bias, respectively. For $\gamma=0$ the swimming direction is biased 
toward $\hat{\bf y}$. The perpendicular component of the
diffusion tensor can be found exactly \cite{bearon2011}
\begin{equation}
D^{(0)}_\perp(\lambda)= \frac{1}{\lambda^2}\left(1-\frac{1}{(I_0(\lambda))^2}\right)\,.
\label{eq:dperp}
\end{equation}
For the parallel one only an approximate expression (not reported
here, as not needed in the following) can be found.

When $\gamma\neq 0$, swimming is biased toward a
  direction that depends on $r$. Our approximation consists in taking
$r$ fixed in Eq.~(\ref{eq2d:thetadot}) and finding the stationary PDF
of $\theta$ which will depend on $r$ parametrically.  This is a sort
of adiabatic approximation, valid when $r$ does not change much in the
time scale over which the PDF of orientation becomes stationary. In
this way the average swimming direction will depend on $r$ as
\begin{equation}
\hat{\bf b}=\left(\!\frac{-\gamma r}{\sqrt{1\!+\!(\gamma r)^2}},\frac{1}{\sqrt{1\!+\!(\gamma r)^2}}\!\right)\,,
\label{eqd2d:fhat}
\end{equation}
being at an angle $-\Phi(r)$ with $\Phi(r)=\mathrm{arctan}(\gamma r)$
with respect to $\hat{\bm y}$.  Since $r$ is assumed fixed, the PDF of
$\theta$ will simply be (\ref{eq2d:pdftheta}) with mean $-\Phi(r)$,
i.e.
\begin{equation}
f^{(\gamma)}(\theta;r)=f^{(0)}(\theta+\Phi(r))=
 \frac{e^{ \lambda \frac{(\cos\theta-\gamma r\sin\theta)}{\sqrt{1+(\gamma r)^2}}}}{2\pi I_0(\lambda)} \,.
\end{equation}

Now the drift and diffusion tensor for the $\gamma\neq 0$ case can be
obtained from those computed at $\gamma=0$ simply changing to a new
(position dependent) frame of reference, essentially we need to rotate
the $\gamma=0$ solution at each point matching the vertical with the
local biasing direction.  Introducing the rotation matrix
$$
\mathbb{R}=
\begin{pmatrix}
\frac{1}{\sqrt{1+(\gamma r)^2}} & -\frac{\gamma r}{\sqrt{1+(\gamma r)^2}} \\
\frac{\gamma r}{\sqrt{1+(\gamma r)^2}}  & \frac{1}{\sqrt{1+(\gamma r)^2}} 
\end{pmatrix}\,,
$$
 the drift can be expressed as
\begin{equation}
\bm V^{(\gamma)} = \mathbb{R} \bm V^{(0)}=
v_s \frac{I_1(\lambda)}{I_0(\lambda)} \, \hat{\bf b}
\end{equation}
with $\hat{\bf b}$ given in (\ref{eqd2d:fhat}), and the diffusion tensor
by $\mathbb{D}^{(\lambda)}= \mathbb{R} \mathbb{D}^{(0)}
\mathbb{R}^T$. As for the latter we are mainly interested in the
radial component which reads
\begin{equation}
D^{(\gamma)}_{rr}= \frac{v_s^2}{d_{\mathrm r}}\frac{D^{(0)}_\perp(\lambda)+(\gamma r)^2 D^{(0)}_\parallel(\lambda)}{1+(\gamma r)^2} \,.\label{eq:dxx-rot2d}
\end{equation}

Now we can approach the advection diffusion equation
(\ref{eq:11}). In particular, because neither the drift nor the
diffusion tensor depends on $z$, we can 
integrate over $z$ and derive the equation for the radial dynamics,
which is the one of interest for us,
\begin{equation}
\partial_t n-\partial_r \left( v_s \frac{I_1(\lambda)}{I_0(\lambda)} \frac{\gamma r \,n}{\sqrt{1+(\gamma r)^2}} + D^{(\gamma)}_{rr} \partial_r n \right) =0\,.
\end{equation}
Imposing  stationarity in the above equation, the resulting
ordinary differential equation can be integrated obtaining
\begin{equation}
n_s(r)= \mathcal{N} \exp(-G(r)) 
\label{eq:gofr}
\end{equation}
with $\mathcal{N}$ a suitable normalizing constant and 
\begin{equation}
 G(r)=\frac{d_{\mathrm r}}{v_s \gamma}  \frac{I_1}{I_0}
\frac{\mathcal{D}}{D^{(0)}_\parallel} (g(r)-\arctan(g(r)))
\end{equation}
with $\mathcal{D}\!=\!\sqrt{(D^{(0)}_\perp\!-\!D^{(0)}_\parallel)/D^{(0)}_\parallel}$
and $g(r)\!=\!\sqrt{1\!+\!(\gamma r)^2}/\mathcal{D}$
In the limit  $(\gamma r)\ll 1$, $G(r)$  can be expanded in
\begin{equation}
G(r)\approx \frac{1}{2} \frac{\gamma d_{\mathrm r} I_1(\lambda)}{v_s D^{(0)}_\perp(\lambda) I_0(\lambda)} \,r^2    = \frac{1}{4} \frac{\gamma}{v_sB F^2_2(\lambda)} r^2\,,
\label{eq:gauss2d}
\end{equation}
with
\begin{equation}
F_2(\lambda)= \left[\frac{I_0(\lambda)-I^{-1}_0(\lambda)}{\lambda I_1(\lambda)}\right]^{1/2}\,.
\label{eq:formula2d}
\end{equation} 
Hence the stationary population takes a Gaussian form.

\begin{figure}[htb!]
\centering
\includegraphics[width=0.9\columnwidth]{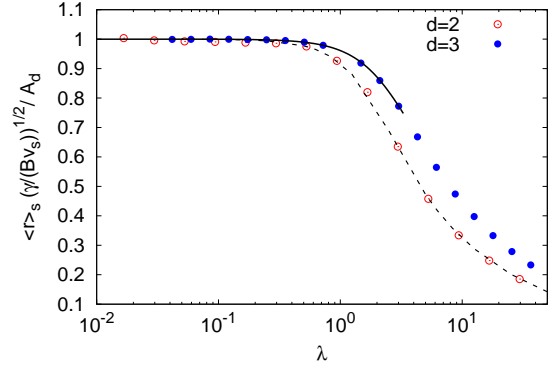}
\caption{Online version in colour.
Comparison between analytical computation of the average
radial distance at stationarity and the numerical computation, in
both two and three dimensions. Normalized average radial distance at
stationarity, $\langle r\rangle_s ({\gamma}/{v_sB})^{1/2}/ A_d$
(with $A_2=\sqrt{4/\pi}$ and $A_3=\sqrt{2/\pi}$) vs $\lambda$
compared with the analytical predictions $F_2(\lambda)$ and
$F_3(\lambda)$ given in Eqs.~(\ref{eq:formula2d}) and
(\ref{eq:formula3d}) respectively (lines). For the case $d=3$
the convergence of the recursion relations requires very high
accuracy for $\lambda>3$.
}
\label{fig:app}
\end{figure}

Then we can compute the average radial distance $\langle r\rangle_s$
at stationarity as
\begin{equation}
\langle r\rangle_s = \frac{\int_0^{\infty} r n_s(r) dr }{\int_0^{\infty} n_s(r) dr}=  \left(\frac{v_sB}{\gamma}\right)^{1/2} \sqrt{\frac{4}{\pi}} F_2(\lambda)\,,
\label{eq:rmed2d}
\end{equation}
in very good agreement with simulations (Fig.~\ref{fig:app}).

\subsection{Analytical approximation in $d=3$\label{suapp:3d}}
The computation in $d=3$ can be performed following step by step the
procedure above described for the two-dimensional model.

For $\gamma=0$ in $d=3$, unlike $d=2$, we do not have closed
expressions. However, there are exact results expressing the
quantities of interest in terms of series in $\lambda=1/(2Bd_r)$,
which have been obtained in Ref.~\cite{bearon2012} (using previous
results from Refs.~\cite{Pedley1987,pk_jfm90,Pedley1992}). We briefly
summarize the results in the following.

For $\gamma=0$, the orientation distribution is the von
Mises-Fisher distribution \cite{mardia}
\begin{equation}
f^{(0)}({\bf p})= \mu(\lambda) e^{\lambda {\bf p}\cdot\hat{\bm z}}= \mu(\lambda) e^{\lambda \cos\theta}\,,
\end{equation}
with ${\bf p}\!\!=\!\!(\cos\phi\sin\theta, \sin\phi\sin\theta, \cos\theta)$ 
and $\mu(\lambda)\!=\!4\pi\sinh\lambda/\lambda$.
The drift has the following exact
expression~\cite{Pedley1987,pk_jfm90,Pedley1992}
\begin{equation}
\bm V^{(0)}=v_s \int d{\bf p} {\bf p}f^{(0)}({\bf p})=v_s (0,0,K_1(\lambda))
\end{equation} 
with $K_1(\lambda)=\mathrm{coth}\lambda-1/\lambda$; while the diffusion
tensor is 
\begin{equation}
\mathbb{D}^{(0)}=\frac{v_s^2}{d_{\mathrm r}}\mathrm{Diag}\{D^{(0)}_\perp(\lambda),D^{(0)}_\perp(\lambda), D^{(0)}_\parallel(\lambda)\}
\end{equation}
with the two entries perpendicular to the direction of gravity equal.
As for the $d=2$ we only need the perpendicular component which takes the
form~\cite{bearon2012}
\begin{equation}
D^{(0)}_\perp(\lambda)={J_1(\lambda)}{\lambda^{-2}}\,,
\end{equation}
with the numerator given by the series
\begin{equation}
J_1(\lambda)= \frac{4\pi}{3} \lambda\mu(\lambda) \sum_{k=0}^{\infty} \lambda^{2k+1} a_{2k+1,1}\,.
\end{equation}
The coefficients $a_{k,n}$ can be obtained by the recursion (see
Refs.~\cite{bearon2012,pk_jfm90} for details):
\begin{eqnarray}
a_{k+1,n} &=& - \frac{(n+2)a_{k,n+1}}{(n+1)(2n+3)} 
+ \frac{(r-1)a_{k,n-1}}{r(2r-1)}+ \frac{b_{k+1,n}}{n(n+1)}\,, \nonumber\\
b_{k+1,n}&=&\frac{(2n+1)\Gamma\left(\frac{n+1}{2}\right)\Gamma\left(\frac{n+2}{2}\right)}{\Gamma(n+1)\Gamma\left(4\frac{k-n+3}{2}\right)\Gamma\left(\frac{k+n+4}{2}\right)} \quad \mathrm{if}\; k+n\; \mathrm{odd}  \nonumber
\end{eqnarray}
and $b_{k+1,n}=0$ otherwise.

With the appropriate rotation matrix, and following the steps described in
Appendix~\ref{suapp:2d}, we obtain the very same expressions found in $d=2$
for the radial component of diffusion tensor (see Eq.~(\ref{eq:dxx-rot2d})).
In particular, we have that at stationarity in the limit $(\gamma r)\ll 1$
the population density takes the expression (\ref{eq:gofr}) with
\begin{equation}
\strut\hspace{-0.2truecm} 
G(r)\!=\! \frac{1}{2} \frac{\gamma r^2}{v_sB F^2_3(\lambda)}
\quad \mathrm{with}\;
F_3(\lambda)\!= \!\left[\frac{2J_1(\lambda)}{\lambda K_1(\lambda)}\right]^{1/2}.
\label{eq:formula3d}
\end{equation} 
Hence, at stationarity, we find the average radial distance 
\begin{equation}
\langle r\rangle_s = \frac{\int_0^{\infty} r n_s(r) dr }{\int_0^{\infty} n_s(r) dr}=  \left(\frac{v_sB}{\gamma}\right)^{1/2} \sqrt{\frac{\pi}{2}} F_3(\lambda)\,.
\end{equation}
in perfect agreement with simulation results (Fig.~\ref{fig:app}).

\bibliography{biblio}

\end{document}